\documentclass[twoside]{article}
\usepackage{amsmath,latexsym,amssymb,verbatim, enumerate, graphicx}
\usepackage[all]{xy}
\usepackage{picture,graphics,epic,eepic}
\usepackage{qic,epsfig}

\usepackage{theorem}

\newtheorem{defi}{Definition}
\newtheorem{prop}[defi]{Proposition}
\newtheorem{lem}[defi]{Lemma}

\newtheorem{theo}[defi]{Theorem}
\newtheorem{coro}[defi]{Corollary}

\newtheorem{probl}{Problem}

\newtheorem{example}{Example}

\newcommand{\ket}[1]{|#1\rangle}
\newcommand{\bra}[1]{\langle#1|}
\newcommand{\braket}[1]{|#1\rangle\langle#1|}

\newcommand \Tr {\mathrm{Tr}}
\newcommand{\nc}{\newcommand}

\nc{\smfrac}[2]{\mbox{$\frac{#1}{#2}$}}

\def\squareforqed{\hbox{\rlap{$\sqcap$}$\sqcup$}}
\def\qed{\ifmmode\squareforqed\else{\unskip\nobreak\hfil
\penalty50\hskip1em\null\nobreak\hfil\squareforqed
\parfillskip=0pt\finalhyphendemerits=0\endgraf}\fi}

\def\>{\rangle}
\def\<{\langle}

\def\>{\rangle}
\def\<{\langle}

\def\be{\begin{equation}}
\def\ee{\end{equation}}
\def\bee{\begin{eqnarray*}}
\def\eee{\end{eqnarray*}}

\begin{document}
\setlength{\textheight}{8.0truein}    

\runninghead{Assisted Entanglement Distillation}
            {N. Dutil and P. Hayden}

\normalsize\textlineskip
\thispagestyle{empty}
\setcounter{page}{1}

\copyrightheading{11}{5\&6}{2011}{0496--0520}

\vspace*{0.88truein}

\alphfootnote

\fpage{1}

\centerline{\bf
ASSISTED ENTANGLEMENT DISTILLATION}
\vspace*{0.37truein}
\centerline{\footnotesize
NICOLAS DUTIL\footnote{ndutil79@gmail.com}}
\vspace*{0.015truein}
\centerline{\footnotesize\it School of Computer Science, Mcgill University, 3480 University Street}
\baselineskip=10pt
\centerline{\footnotesize\it Montr\'eal, Qu\'ebec, H3A 2A7, Canada}
\vspace*{10pt}
\centerline{\footnotesize
PATRICK HAYDEN \footnote{patrick@cs.mcgill.ca}}
\vspace*{0.015truein}
\centerline{\footnotesize\it School of Computer Science, Mcgill University, 3480 University Street}
\baselineskip=10pt
\centerline{\footnotesize\it Montr\'eal, Qu\'ebec, H3A 2A7, Canada}
\vspace*{0.015truein}
\centerline{\footnotesize\it Perimeter Institute for Theoretical Physics,
    31 Caroline St. N.}
\baselineskip=10pt
\centerline{\footnotesize\it Waterloo, Ontario,
    N2L 2Y5, Canada}
\vspace*{0.225truein}

\publisher{November 15, 2010}{April 7, 2011}

\vspace*{0.21truein}

\abstracts{
Motivated by the problem of designing quantum repeaters, we study entanglement distillation between two parties, Alice and Bob, starting from a mixed state and with the help of ``repeater'' stations. To treat the case of a single repeater, we extend the notion of entanglement of assistance to arbitrary mixed tripartite states and exhibit a protocol, based on a random coding strategy, for extracting pure entanglement. The rates achievable by this protocol formally resemble those achievable if the repeater station could merge its state to one of Alice and Bob even when such merging is impossible. This rate is provably better than the hashing bound for sufficiently pure tripartite states. We also compare our assisted distillation protocol to a hierarchical strategy consisting of entanglement distillation followed by entanglement swapping. We demonstrate
by the use of a simple example that our random measurement strategy outperforms hierarchical
distillation strategies when the individual helper stations' states fail to individually factorize into 
portions associated specifically with Alice and Bob.
Finally, we use these results to find achievable rates for the more general scenario, where many spatially separated repeaters help two recipients distill entanglement.
}{}{}

\vspace*{10pt}

\vspace*{3pt}

\vspace*{1pt}\textlineskip    

\section{Introduction} \label{sec:introduction}

\noindent
Establishing entanglement over long distances is very difficult due to the cumulative effects of noise, which lead to an exponential degradation of entanglement fidelity with distance. Establishing high-quality entanglement, however, is a necessary prerequisite for performing quantum teleportation~\cite{teleportation} between far distant laboratories or developing device-independent cryptographic technologies~\cite{ekert,device-indep}. One strategy for dealing with this difficulty is to employ quantum repeaters, stations intermediate between the sender and receiver that can participate in the process of entanglement distillation, thereby improving on what the sender and receiver could do on their own~\cite{repeaters,hierarchy,repeat-exp1,repeat-exp2}. In this article, we introduce and study a version of the entanglement of assistance~\cite{dfm} appropriate to the study of such mixed state assisted distillation problems in the context of quantum Shannon theory.

A single copy version of this problem was analyzed in the context of spin chains under the name of localizable
entanglement~\cite{localize}. This quantity, along with others such as the average singlet conversion
probability (SCP)~\cite{cirac,vidal,entangleDistr}, can be used as figures of merit for characterizing quantum networks.
A quantum network~\cite{cirac,entangleDistr,statetransfer,functional,cirac02} consists of spatially
separated nodes connected by quantum communication
channels. Each node of the network represents local physical systems which hold quantum information, stored in quantum memories. The information stored at the node can then be processed locally by using optical beam splitters~\cite{Beam} and planar lightwave circuits~\cite{Lightwave}, among other technologies. Entanglement between neighboring nodes can be established by locally preparing a state at one node and distributing part of it to the neighboring node using the physical medium connecting the two nodes. One of the main tasks then becomes the design of protocols that use the entanglement between the neighboring nodes to establish pure entanglement between the non adjacent nodes.
\begin{figure} [t]
\centerline{\epsfig{file=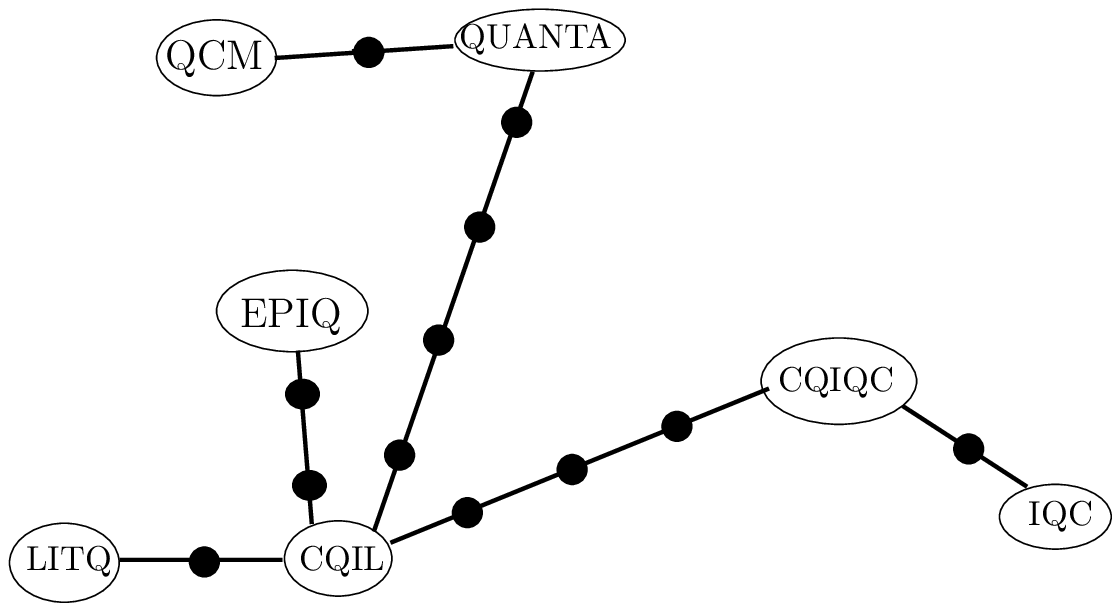, width=10.2cm}} 
\vspace*{13pt}
\fcaption{\label{fig:network1}A hypothetical quantum network connecting various university quantum laboratories. Repeater stations are represented by black dots.}
\end{figure}

In~\cite{cirac,entangleDistr}, several strategies were analyzed for one-dimensional and two-dimensional geometries. In these simplified networks, each pair of nodes shares a pure state $\ket{\psi}=\sqrt{\lambda_1}\ket{00} + \sqrt{\lambda_2}\ket{11}$. For the case of a one-dimensional chain with only one repeater node, it was found that by applying an entanglement swapping protocol consisting of a Bell measurement at the repeater node, one can achieve a singlet conversion probability equal to the optimal singlet conversion probability for the state $\ket{\psi}$. That is, the repeater node does not reduce the likelihood of recovering a maximally entangled pair between Alice and Bob's laboratories. This desirable property is not preserved, however, for chains consisting of many repeater stations separating the two laboratories. It was shown in~\cite{cirac} that no measurement strategy can keep the SCP between Alice and Bob from decreasing exponentially with the number of repeaters, making them useless for establishing entanglement over long distances.

One way to deal with this problem is to introduce redundancy in the network~\cite{repeaters}. By preparing and distributing many copies of the state $\ket{\psi}$ across the chain, the repeater stations will be able to help Alice and Bob in producing singlets. The redundancy introduced in the network allows the stations to perform joint measurements on their shares, concentrating the entanglement found in each copy of $\ket{\psi}$ into a small number of highly entangled particles. For one-dimensional chains, the rate at which entanglement can be established between the two endpoints will approach the entropy of entanglement $S(A)_{\psi}$, no matter the number of repeaters introduced between the endpoints. The more copies of the state $\ket{\psi}$ are prepared and distributed between the nodes, the more transparent the repeaters will become, allowing us to view the entire chain as a noiseless channel for Alice and Bob.

This is an ideal situation unlikely to occur in real experiments, as only a finite number of copies of the state $\ket{\psi}$ will be prepared and the preparation and distribution of copies of this state across the network will be imperfect. It is also reasonable to assume that the storage of many qubits at a repeater station, or at one of the laboratories, will be more prone to errors over time than the storage of a single qubit. Hence, the global state of a quantum network will most likely be mixed. For such mixed state networks, we can ask the question: how much entanglement can we establish between Alice and Bob by performing local operations and classical communication (LOCC) on the systems part of the network ?

In this paper, we extend the models previously studied in~\cite{cirac, entangleDistr} to allow for an arbitrary mixed state between adjacent nodes. This is an initial step towards handling more complex and realistic situations. First, we will consider a network consisting of two receiving nodes (Alice and Bob), separated by a repeater node (Charlie), whose global state is a mixed state $\psi^{ABC}$. We study the optimal distillable rate achievable for Alice and Bob when assistance from Charlie is available. This problem reduces to the two-way distillable entanglement for states in a product form $\psi^{C} \otimes \psi^{AB}$. There is currently no simple formula for computing the two-way distillable entanglement of a bipartite state $\psi^{AB}$, which has been studied extensively by Bennett et al. and others in~\cite{Bennett,PurNoisy,interpolation,adaptive}. We do not attempt to solve this problem here, and turn our attention instead to good computable lower bounds for assisted distillation of mixed states. We provide a bound which generically exceeds the hashing inequality for states $\psi^{ABC}$ when the coherent information from $C$ to $AB$ is positive. We will also consider a more general scenario where many spatially separated helpers perform measurements on their share of the state and send their results to two recipients, who then exploit this information to extract a greater amount of maximally entangled pairs.

\textbf{Structure of the paper:} Section \ref{sec:task} begins with a quick review of the main results for the entanglement of assistance problem, and introduces the extension to the general mixed state scenario. In Section \ref{sec:EoA}, we define the one-shot entanglement of assistance. We show that its regularization is equal to the optimal assisted distillation rate achievable for the scenario introduced in Section \ref{sec:task}. We then derive two upper bounds for this one-shot quantity, and give some examples of classes of states attaining them. Next, in Section \ref{sec:coding}, we introduce a protocol, based on a random coding strategy, which achieves a rate that is strictly better than the hashing bound under appropriate conditions.
Finally, in Section \ref{sec:multi}, we generalize the previous scenario to the case of many helpers. An appendix, containing a few technical proofs, appears at the end.

\textbf{Notation:} In this paper, we restrict our attention to finite dimensional Hilbert spaces. Quantum systems under consideration will be denoted $A, B,\ldots,$ and are freely associated with their Hilbert spaces, whose (finite) dimensions are denoted $d_A, d_B, \ldots$. If $A$ and $B$ are two Hilbert spaces, we write $AB \equiv A \otimes B$ for their tensor product. Unless otherwise stated, a "state" can be either pure or mixed. The symbol for such a state (such as $\psi$ and $\rho$) also denotes its density matrix. The density matrix $\braket{\psi}$ of a pure state will frequently be written as $\psi$. We write $S(A)_{\psi} = -\Tr( \psi^A \log \psi^A)$ to denote the von Neumann entropy of a density matrix $\psi^A$ for the system $A$. The function $F(\rho,\sigma) := \Tr \sqrt{\rho^{1/2}\sigma\rho^{1/2}}$ is the Uhlmann fidelity between the two states $\rho$ and $\sigma$~\cite{uhlmann:fid,jozsa:fid}. The trace norm $\|X\|_1$ of an operator $X$ is defined to be $\Tr|X| = \Tr \sqrt{X^{\dag}X}$. The typical subspace $\tilde{A}^n_{\psi,\delta}$ associated with the state $\psi^{\otimes n}_A$ is written as $\tilde{A}$. Finally, the probability of an event $X$ is denoted as $P(X)$.

\section{The Task} \label{sec:task}
\noindent
Refs.~\cite{dfm} and~\cite{Cohen2} introduced the following quantity, called the \emph{entanglement of assistance} of a bipartite mixed state $\rho^{AB}$:
\begin{equation}
\label{eq:EofA} E_A(\rho^{AB})
      := \max_{{\cal E}}  \sum_i p_i S(\psi_i^{AB}),
\end{equation}
where the maximum is over all decompositions of $\rho^{AB}$ into
a convex combination of pure states ${\cal E} = \{p_i,
\psi_i^{AB}\}$. Suppose that $\ket{\psi}^{ABC}$ is a purification
of $\psi^{AB}$. By acting on the purification system $C$, the
helper Charlie can effect any such pure state convex decomposition
$\rho^{AB} = \sum_i p_i\psi_i^{AB}$ for Alice and Bob's
state~\cite{schroedinger,Hughston}, and so the quantity $E_A$
maximizes the amount of entanglement that Alice and Bob can
distill with help from Charlie. Since $E_A$ is not, in general,
additive under tensor products~\cite{dfm}, it will often be the
case that entangled measurements performed by Charlie on the joint
state $(\psi^{C})^{\otimes n}$ will be more beneficial to Alice
and Bob than separate measurements on individual copies of
$\psi^{C}$. Thus, allowing for many copies of the state
$\psi^{ABC}$, we can ask: \emph{what is the optimal asymptotic
distillable rate between Alice and Bob with help from Charlie
under \mbox{LOCC?}} The answer to this question was given in~\cite{SVW}, where it was shown that the optimal asymptotic rate,
denoted by $E_A^{\infty}$, is equal to the regularization of
$E_A$: \be E_A^{\infty}(\psi^{ABC})=\lim_{n \rightarrow \infty}
\frac{1}{n} E_A(\psi_{ABC}^{\otimes n}). \ee  Furthermore, a
simple expression for $E_A^{\infty}$ in terms of entropic
quantities was also obtained:
\begin{equation}
\label{thm:EofA}
E_A^{\infty}(\psi^{ABC})=\min\{S(A)_\psi,S(B)_\psi\}.
\end{equation}
A detailed proof of this statement can be found in~\cite{SVW}. To understand how powerful third-party assistance is, we need only observe that if $\psi^{AB}$ were pure, the distillable entanglement would be given by the entropy $S(A)_\psi = S(B)_\psi$. Thus, assistance by a third party holding the purification $C$ is equivalent to giving $C$ to one of Alice and Bob, whichever will result in the least bipartite entanglement. (The achievability of these rates will also follow from our more general Theorem~\ref{thm:lowerbound}.)

\begin{figure} [t]
\centerline{\epsfig{file=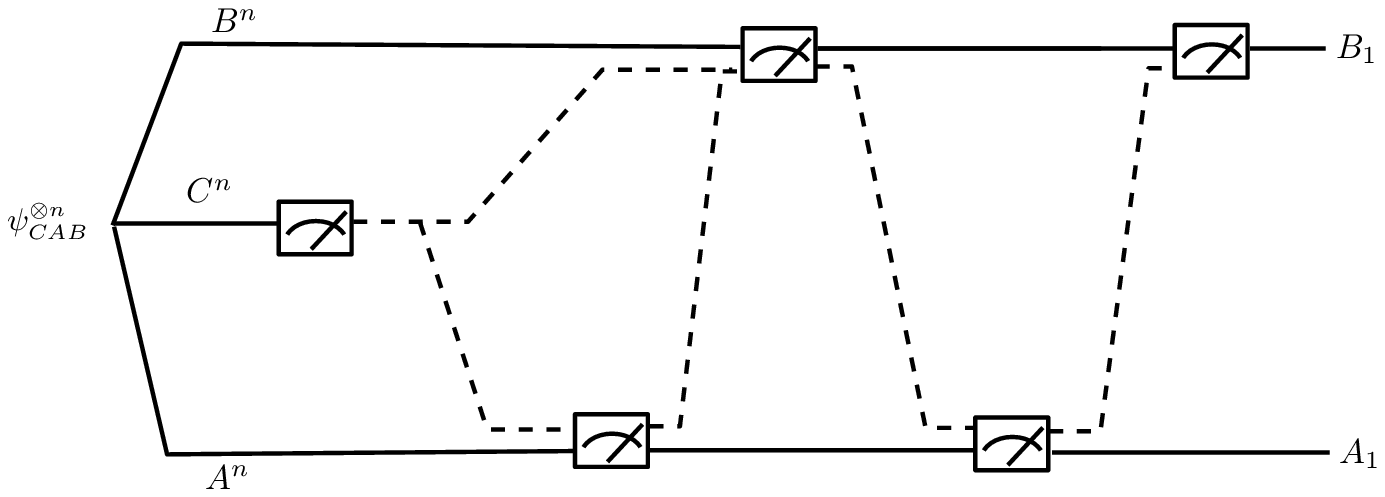, width=13.2cm}} 
\vspace*{13pt}
\fcaption{\label{fig:b1s}Quantum circuit representing a broadcast assisted entanglement distillation protocol. Solid lines indicate quantum information and dashed lines classical information. Charlie first performs a measurement, sending copies of the classical outcome to Alice and Bob. Alice and Bob then implement an LOCC operation, conditioned on that classical outcome.}
\end{figure}

In this section, we extend these ideas to the case of a general mixed state, with a measurement of
Charlie's system followed by an entanglement distillation protocol between Alice and Bob. The problem is illustrated in Figure~\ref{fig:b1s}.

\begin{probl}[\textbf{Broadcast, Assisted Distillation}] \label{pbl:oneway}
Given many copies of a tripartite \textit{mixed} state $\psi^{ABC}$ shared between two recipients (Alice and Bob) and a helper (Charlie), find the optimal rate of entanglement distillable between Alice and Bob with the help of Charlie if no feedback communication is allowed: Charlie performs a positive operator valued measure (POVM) and broadcasts the measurement outcome to Alice and Bob, who proceed to distill. The optimal rate is denoted by $D_A^{\infty}(\psi^{ABC})$. It is the asymptotic entanglement of assistance.
\end{probl}
We call a protocol which satisfies the constraint of Problem \ref{pbl:oneway} a \emph{broadcast assisted distillation protocol}. More
formally, it consists of
\newcommand{\LOCC}{{\cal V}_x}
\begin{romanlist}
 \item A POVM $E=(E_x)_{x=1}^X$ for Charlie. Without loss of generality, we can assume that the operators $E_x$ are all of rank one.
 \item For each $x$, an LOCC operation $\LOCC: A^nB^n \rightarrow A_1B_1$, where $A_1$ and $B_1$ are subspaces of $A^n$ and $B^n$ of equal dimensions, implemented by Alice and Bob.
\end{romanlist}
We refer to a broadcast assisted protocol as an $(n,\epsilon)$-protocol if it acts on
$n$ copies of the state $\psi^{ABC}$ and produces a maximally
entangled state of dimension $M_n:=d_{A_1}$
\begin{equation}
\ket{\Phi^{M_n}} = \frac{1}{\sqrt{{M_n}}}
\sum_{m=1}^{M_n} \ket{m}^{A_1} \otimes\ket{m}^{B_1}
\end{equation}
up to fidelity $1- \epsilon$:
\begin{equation} F^2 \biggr (\Phi^{M_n}, \sum_{x=1}^X p(x) \LOCC(\psi_x^{A^nB^n}) \biggl )
\geq 1-\epsilon,
\end{equation}
where
\begin{equation}
\psi_x^{A^nB^n} = \frac{1}{\Tr_{C^n}[E_x(\psi^C)^{\otimes
n}]}\Tr_{C^n} \biggr [ (E_x \otimes I^{AB}) (\psi^{ABC})^{\otimes
n}  \biggl ].
\end{equation}

A real number $R \geq 0$ is said to be an achievable rate if there exists, for every $n$ sufficiently large, an
$(n,\epsilon)$-protocol with $\epsilon \rightarrow 0$ and $\frac{1}{n} \log{M_n} \rightarrow R$ as $n \rightarrow
\infty$. Lastly, we have
\begin{equation}
D_A^{\infty}(\psi^{ABC}) := \sup \{R: R \text{ is achievable} \}.
\end{equation}

The restriction to POVMs with rank one operators in the preceding definition can be justified as follows: any POVM $F$ containing positive operators with rank higher than one that Charlie would wish to perform can be simulated by a POVM $E$ with rank one operators on Charlie's system followed by some processing by Alice and Bob. More precisely, suppose Charlie wants to perform a POVM $F=\{F_x\}$ on his state with some operators having rank greater than one. Consider the spectral decomposition of each operator:
\begin{equation}
F_x = \sum_i \lambda^x_i \braket{\alpha^x_i},
\end{equation}
where $\{\ket{\alpha^x_i}\}$ are eigenvectors of $F_x$ with eigenvalues $\{\lambda^x_i\}$. Then $E = \{\lambda^x_i\braket{\alpha^x_i}\}_{x,i}$ is a POVM with rank one operators. Instead of performing the POVM $F$, Charlie does a measurement corresponding to the POVM $E$. After Alice and Bob receive the measurement outcome, the state is given by
\begin{equation}
\psi^{AA_1A_2BB_1B_2} = \sum_{x,i} q_{x,i} \psi^{AB}_{x,i} \otimes \braket{xx}^{A_1B_1} \otimes \braket{ii}^{A_2B_2}.
\end{equation}
To simulate $F$ being performed by Charlie, Alice and Bob can trace out the $A_2$ and $B_2$ systems. The state becomes
\begin{equation}
\psi^{AA_1BB_1} = \sum_{x} p_x \psi^{AB}_{x} \otimes \braket{xx}^{A_1B_1},
\end{equation}
with $\psi^{AB}_x := \smfrac{1}{p_x}\sum_i q_{x,i} \psi^{AB}_{x,i}$ and $p_x = \sum_i q_{x,i}$. Observe that this preprocessing can be embedded within the LOCC operation $\LOCC$. Hence, there is no loss of generality in assuming POVMs with rank one operators in step 1 of the protocol.

For pure states, $D^{\infty}_A(\psi)$ reduces to the asymptotic entanglement of assistance $E^{\infty}_A(\psi)$. For product states of the form $\psi^{C} \otimes \psi^{AB}$, $D^{\infty}_A(\psi)$ is equivalent to the two-way distillable entanglement $D(\psi^{AB})$. A formula is known for the two-way distillable entanglement (see Theorem 15 in Devetak and Winter~\cite{DW}), but its calculation is intractable for most states. We will instead use the hashing bound to the one-way distillable entanglement $D_{\rightarrow}(\psi^{AB})$~\cite{DW}, in which only communication from Alice to Bob is permitted, and which is more tractable. We remind the reader of the result for convenience:

\begin{lem}[Hashing inequality~\cite{Bennett,DW}]
\label{hashinginequality} Let $\psi^{AB}$ be an arbitrary bipartite mixed state. Then,
\begin{equation}\label{eq:hashing}
D_{\rightarrow}(\psi^{AB}) \geq S(B)_{\psi} - S(AB)_{\psi} =: I(A\rangle B)_{\psi}.
\end{equation}
\end{lem}

\section{One-shot Entanglement of Assistance} \label{sec:EoA}

\subsection{Definition and connection to assisted distillation} \label{subsec:defns}
\noindent
As mentioned before, it was shown in~\cite{SVW} that for pure states, the operationally defined quantity $E_A^{\infty}$ corresponds to the regularization of the one-shot entanglement of assistance $E_A$. In a similar fashion, we define the
\emph{one-shot entanglement of assistance} $D_A(\psi^{ABC})$ of a tripartite mixed state $\psi^{ABC}$ and show
that its regularization is equal to $D^{\infty}_A(\psi^{ABC})$. We then look at some of the properties of $D_A(\psi^{ABC})$.

\begin{defi}  \label{def:A1D}
For an arbitrary state $\psi^{ABC}$, define
\begin{equation}
 \begin{split}\label{eq:A1D}
  D_A(\psi^{ABC}) &:= \sup_{\substack{E=\{E_x\}}} \left\{ \sum_x p_x D(\psi_x^{AB}) \bigg | \psi_x^{AB} = \frac{1}{p_x} \mathrm{Tr}_C[(E_x \otimes I_{AB})\psi^{ABC}] \right\},
 \end{split}
\end{equation}
where $p_x = \mathrm{Tr}[E_x \psi^C]$ and the supremum is taken over all POVMs $E=\{E_x\}$ with rank one operators on Charlie's system $C$.
\end{defi}

The quantity $D_A(\psi^{ABC})$ can also be characterized using a maximization over all pure state decompositions $\{p_i,
\psi_i^{ABR}\}$ of the purified state $\psi^{ABCR}$:
\begin{prop}\label{prop:equiv}
Let $\psi^{ABC}$ be an arbitrary state, with purification $\psi^{ABCR}$, then
  \begin{equation}
   \label{eq:equiv}
     D_A(\psi^{ABC}) = \sup_{ \{p_i, \psi_i^{ABR}\}}   \sum_i p_i D(\psi_i^{AB}),
 \end{equation}
where the supremum is taken over all ensembles of pure states
$\{p_i, \psi_i^{ABR}\}f$ satisfying $\sum_i p_i \psi_i^{ABR} =
\mathrm{Tr}_C \psi^{ABCR}$.
\end{prop}
{\bf Proof:} Any rank one POVM on $C$ induces an ensemble of pure states on $ABR$ with average state $\psi^{ABR}$ and for every such ensemble there exists a corresponding POVM~\cite{Hughston}. Applying this observation to the definition of the one-shot entanglement of assistance yields the result. \square\,

We can interpret Eq.~(\ref{eq:equiv}) as follows: by varying a
POVM on his state, Charlie can collapse the purified state
$\psi^{ABCR}$ into any pure state ensemble decomposition
$\{p_i,\psi_i^{ABR}\}$ for the $A$,$B$, and $R$ systems. Since we
don't have access to the purifying system $R$, the quantity $D_A(\psi^{ABC})$
maximizes the average amount of distillable entanglement
between Alice and Bob. The next result shows that the regularized
version of $D_A$ is in fact equal to the asymptotic entanglement of assistance $D_A^{\infty}(\psi^{ABC})$.

\subsection{Basic properties} \label{subsec:properties}

\begin{theo}[Equivalence]
\label{thm:equivalence} Let $\psi^{ABC}$ be an arbitrary tripartite state. Then the following equality holds:
\begin{equation}\label{eq:ensEq}
D_A^{\infty}(\psi^{ABC}) = \lim_{n \rightarrow \infty} \frac{1}{n} D_A\biggl ((\psi^{ABC})^{\otimes n} \biggr ).
\end{equation}
\end{theo}
{\bf Proof:} We demonstrate the ``$\leq$'' first. Consider any achievable rate $R$ for a broadcast assisted distillation protocol. By definition, there exists, for every $n$ sufficiently large, an $(n, \epsilon)$-protocol with $\epsilon \rightarrow 0$ and $\smfrac{1}{n}\log(M_n)\rightarrow R$ as $n \rightarrow \infty$. For a protocol working on $n$ copies of the state $\psi^{ABC}$, denote Charlie's POVM by $E=(E_x)_{x=1}^X$, and for each outcome $x$, the LOCC operation implemented by Alice and Bob by $\LOCC$. Write
\begin{align}
\Omega^{A_1B_1} &:= \sum_{x=1}^X p_x \LOCC (\psi_x^{A^nB^n}) \notag \\
&= \sum_{x=1}^X p_x \Omega_x^{A_1B_1},
\end{align} where $p_x =\mathrm{Tr}[E_x (\psi^{C})^{\otimes n}]$ and $\psi_x^{A^nB^n} =\frac{1}{p_x} \Tr_{C^n}[ (E_x \otimes I_{AB})\psi_{ABC}^{\otimes n}]$. The state $\Omega_x^{A_1B_1}$ is the output state of $\LOCC(\psi_x^{A^nB^n})$. By hypothesis, we have
\begin{equation}
F^2(\Phi^{M_n}, \Omega^{A_1B_1}) \geq 1-\epsilon,
\end{equation}
which, shifting to the trace norm, implies
\begin{equation}
\label{eq:dist1} \biggr \|
\Phi^{M_n} - \Omega^{A_1B_1} \biggl \|_1 \leq 2\sqrt{\epsilon} := \epsilon'.
\end{equation}
The trace distance is non-increasing under the partial trace, and so tracing out the $A_1$ system, we have
\begin{equation}
\label{eq:dist2} \biggr \|
\Phi^{M_n}_{B_1} - \Omega^{B_1} \biggl \|_1 \leq \epsilon',
\end{equation}
where $\Phi^{M_n}_{B_1} = \frac{1}{M_n}\sum^{M_n}_{m=1} \braket{m}^{B_1}$.

We can apply the Fannes inequality (Lemma \ref{lem:Fannes}) on Eqs. (\ref{eq:dist1}) and (\ref{eq:dist2}) to get a bound on $\log(M_n)$ in terms of the coherent information of the state $\Omega_{A_1B_1}$:
  \begin{align}
      \log{M_n} &\leq S(B_1)_{\Omega} - S(A_1B_1)_{\Omega} + 3\log(M_n)\eta(\epsilon') \notag \\
      &= I(A_1 \rangle B_1)_{\Omega} +3\log(M_n)\eta(\epsilon'),
  \end{align}
where $\eta(\epsilon')$ is a function which converges to zero for sufficiently small $\epsilon'$. (The definition of $\eta(\epsilon')$ can be found in Lemma $\ref{lem:Fannes}$.)
Using the convexity of the coherent information~\cite{bns}, the hashing inequality, and the definitions of $D$ and $D_A$, we get the following series of inequalities:
\begin{align}
      \log{M_n}&\leq I(A_1\rangle B_1)_{\Omega} + 3\log(M_n)\eta(\epsilon')\notag\\
      &\leq \sum_x p_x I(A_1\rangle B_1)_{\Omega_x} + 3\log(M_n)\eta(\epsilon') \notag\\
      &\leq \sum_x p_x D(\Omega^{A_1B_1}_x) + 3\log(M_n)\eta(\epsilon') \notag\\
      &\leq \sum_x p_x D(\psi_x^{A^nB^n}) +3\log(M_n)\eta(\epsilon')  \notag \\
      &\leq D_A((\psi^{ABC})^{\otimes n}) + 3n\log(d_A)\eta(\epsilon').
  \end{align}
Since $\epsilon \rightarrow 0$ and $\smfrac{1}{n}\log(M_n) \rightarrow R$ as $n\rightarrow \infty$, the achievable rate $R$ is at most $\lim \limits_{\substack{n \to \infty}}\smfrac{1}{n}D_A(\psi^{\otimes n})$, which proves the ``$\leq$'' part since $R$ was arbitrarily chosen.

To show the ``$\geq$'' part, suppose Charlie performs any POVM $E=(E_x)$ on one copy of the state $\psi^{ABC}$ and broadcasts the
result to Alice and Bob. They now share the state
\begin{equation} \tilde{\psi}^{A'ABB'} = \sum_x p_x \braket{x}^{A'} \otimes
\psi_x^{AB} \otimes \braket{x}^{B'}.
\end{equation} Since Alice and Bob know the outcome of Charlie's POVM, the distillable entanglement of $\tilde{\psi}^{A'ABB'}$ is at least
\begin{equation}
D(\tilde{\psi}^{A'ABB'}) \geq \sum_x p_x
D(\psi_x^{AB}).
\end{equation} To see this, consider many copies of ${\psi}^{A'ABB'}$ and let Alice and Bob perform projective measurements on the systems $A'$ and $B'$ for each copy of the state. Group the outcome states into blocks, where each block corresponds to a specific measurement outcome. For each of these blocks, there exist LOCC operations $\LOCC$ which will distill arbitrarily close to the rate $D(\psi^{AB}_x)$. Thus, there is a protocol achieving the rate $\sum_x p_x D(\psi_x^{AB})$, which proves the ``$\geq$'' part. \square\,

Finding a formula for the one-shot quantity $D_A(\psi^{ABC})$ appears to be a difficult problem, and so we look for upper bounds which are attained for a subset of all possible states. For the remainder of this section, we look at two upper bounds and give examples of states attaining them.

\begin{prop}
\label{thm:upperbound} Let $\psi^{ABC}$ be an arbitrary tripartite state. We have the following upper bound for
$D_A(\psi^{ABC})$:
\begin{equation}
D_A(\psi^{ABC}) \leq \inf_{\cal E} \sum_i p_i E_A (\psi^{ABC}_i) ,
\end{equation}
where the infimum is taken over all ensembles of pure states
$\{p_i,\psi^{ABC}_i\}$ such that $\psi^{ABC} = \sum_i p_i
\psi_i^{ABC} $.
\end{prop}

{\bf Proof:} Let $\psi^{ABC} = \sum_i p_i \psi_i^{ABC}$, where the states $\psi_i^{ABC}$ are pure. Consider the following classical-quantum
state $\phi^{ABCX} = \sum_i p_i \psi_i^{ABC} \otimes \braket{i}^X$. If Charlie is in possession of the $X$
system, then
\begin{equation}
D_A (\psi^{ABC}) = D_A(\sum_i p_i \psi_i^{ABC}) \leq D_A(\phi^{ABCX})
\end{equation}
by the definition of $D_A$.
Now, for a pure state $\psi_i^{ABC}$, the
one-shot quantity $D_A(\psi_i^{ABC})$ is also equal to the one-shot entanglement of assistance
$E_A(\psi_i^{ABC})$. For the state $\phi^{ABCX}$, we have
\begin{equation}
    D_A(\phi^{ABCX})
    = \sum_i p_i D_A(\psi_i^{ABC})
    = \sum_i p_i E_A(\psi_i^{ABC}).
\end{equation}
Achievability is obtained by considering the POVM $G=\{G_{ix}\}$ with positive
operators $G_{ix} = E^i_x \otimes \braket{i}^{X}$, where $E^i = \{E^i_{x}\}$ is the POVM maximizing Eq.~(\ref{eq:A1D}) for the state $\psi_i^{ABC}$. Optimality follows from the convexity of $D_A(\phi^{ABCX})$ on the ensemble $\{p_i, \psi^{ABC}_i \otimes \braket{i}^X \}$ (see Proposition \ref{lemma:convex}) and the fact that $D_A(\psi_i^{ABC} \otimes \braket{i}^X) = D_A(\psi^{ABC}_i)$.

Hence, we have $D_A(\psi^{ABC}) \leq \sum_i p_i E_A(\psi_i^{ABC})$, and since this holds for any pure state ensemble $\{p_i, \psi_i^{ABC}\}$, we arrive at the statement of the proposition. \square\,

With this result in hand, we now exhibit a set of states for which we can compute the value of $D_A$ exactly.
\begin{example}
Consider the following family of classical-quantum states, with classical system $C$:
\begin{equation}\psi^{ABC} = \sum_{i=1}^{d_C} p_i \psi_i^{AB} \otimes \ket{i} \bra{i}^C,\end{equation} where
$\psi_i^{AB}$ are pure states. Since $D_A$ is convex on pure ensembles $\{p_i,\psi^{ABC}_i\}$, the quantity
$D_A(\psi^{ABC})$ is upper bounded by $\sum_i p_i D_A(\psi_i^{AB} \otimes \braket{i}^C)$. Since assistance is not helpful for a product state $\psi^{AB} \otimes \phi^{C}$, we have that $D_A(\psi_i^{AB}
\otimes \braket{i}^C) = D(\psi_i^{AB}) = S(A)_{\psi_i}$. By considering the POVM $E=\{\braket{i}^C\}_{i=1}^{d_C}$, we also have $D_A(\psi^{ABC}) \geq \sum_i p_i D(\psi_i^{AB}) = \sum_i p_i S(A)_{\psi_i}$. Hence, for this special class of classical-quantum states, the upper bound is attained and $D_A$ is just the average entropy of the $A$ system for the ensemble $\{p_i,\psi_i^{AB}\}$.
\end{example}

\begin{prop}
\label{thm:upperboundEofA} Let $\psi^{ABC}$ be an arbitrary tripartite state. Then
 \begin{equation}
   D_A(\psi^{ABC}) \leq E_A(\psi^{ABC}).
 \end{equation}
\end{prop}
{\bf Proof:} From Proposition \ref{thm:upperbound} and the concavity of the entanglement of assistance quantity $E_A$ (see~\cite{dfm} for a proof), we have
\begin{align}
  D_A(\psi^{ABC}) &\leq \inf_{{\cal E}} \sum_i p_i E_A(\psi_i^{ABC})\notag  \\
     & \leq \inf_{{\cal E}} E_A(\sum_i p_i \psi_i^{ABC}) \notag \\
     & = E_A(\psi^{ABC})
\end{align}
where the infimum is taken over all pure state ensembles $\{p_i,
\psi_i^{ABC}\}$ of the state $\psi^{ABC}$. \square\,

The previous bound on $D_A$ is better understood by imagining the following scenario. The $A'$ system of a pure state $\psi^{AA'}$ is sent to a receiver (i.e Bob) via a noisy channel ${\cal N}$, which can be expressed in its Stinespring form as ${\cal N}(\psi) = \Tr_E U\rho U^{\dag}$, where $U: A' \rightarrow BE$ is an isometry. Another player, Charlie,  tries to help Alice and Bob by measuring the environment and sending its measurement outcome to Alice and Bob. Two cases can occur. If Charlie has complete access to the environment, the best rate Alice and Bob can achieve is given by the entanglement of assistance $E_A(\psi^{ABC})$. More likely, however, is the case where Charlie will only be able to measure a subsystem $C_1$ of the environment $E=C_1C_2$. In this situation, the optimal rate is given by the one-shot entanglement of assistance $D_A(\psi^{ABC_1})$, where $\psi^{ABC_1} = \Tr_{C_2} \psi^{ABE}$. Since this case is more restrictive to Charlie in terms of measuring possibilities, it makes sense that $D_A(\psi^{ABC}) \leq E_A(\psi^{ABC})$ for any tripartite mixed state $\psi^{ABC}$.  This bound will be attained for all pure states $\psi^{ABC}$ since $D_A$ reduces to $E_A$ in this case.

\section{Achievable Rates for Assisted Distillation} \label{sec:coding}
\noindent
In this section, we find the rates achieved by a random coding
strategy for assisted entanglement distillation. The helper Charlie will simply perform
a random measurement in his typical subspace. In light
of the equivalence demonstrated in the previous section, Eq.~(\ref{eq:ensEq}), we will prove a lower bound on the asymptotic entanglement of assistance by bounding the regularized entanglement of assistance quantity. We will need the following lemma, which is derived from a proposition used in~\cite{merge} (see also Appendix A) in the context of assisted distillation of pure states.
\begin{lem}\label{lem:unionbound}
Suppose we have $n$ copies of the pure state $\psi^{CABR}$ with $S(R)_{\psi} < S(AB)_{\psi}$ and $S(B)_{\psi} < S(AR)_{\psi}$. Let $\psi^{\tilde{C}A^nB^nR^n}$ be be the normalized state obtained after projecting the space $C^n$ into its typical subspace $\tilde{C}$. If Charlie performs a (rank one) random measurement of his system $\tilde{C}$, we have, for any fixed $\xi_1 > 0$ and $\xi_2 > 0$,
\begin{equation}\label{eq:fanness}
\int_{\mathbb{U}(\tilde{C})} P\left (\| \psi^{R^n}_J - (\psi^R)^{\otimes n} \|_1 < \xi_1 \bigcap \| \psi^{B^n}_J - (\psi^B)^{\otimes n}\|_1 < \xi_2\right ) dU \geq 1-\alpha,
\end{equation}
 where $\alpha$ can be made arbitrarily small by taking sufficiently large values of $n$. Here, $J$ is the random variable associated with the measurement outcome and $\psi_J^{A^nB^nR^n}$ is the pure state of the systems $A^n, B^n$ and $R^n$ after Charlie's measurement.
\end{lem}
See Appendix A for a proof of Lemma \ref{lem:unionbound}. The following theorem generalizes the reasoning used in~\cite{merge} to the mixed state case. The lower bound on the rate at which ebits are distilled, involving the minimum of $I(AC\rangle B)_\psi$ and $I(A\rangle BC)_\psi$, suggests that $C$ is merged either to Alice or Bob, at which point they engage in an entanglement distillation protocol achieving the hashing bound. This need not be the case, however. In the discussion following the proof of the theorem, we will exhibit an example where merging is impossible but the rates are nonetheless achieved.

\begin{theo}
\label{thm:lowerbound} Let $\psi^{ABC}$ be an arbitrary tripartite state shared by two recipients (Alice and Bob) and a helper (Charlie). Then the asymptotic entanglement of
assistance is bounded below as follows:
\begin{equation}\label{eq:L}
D_A^{\infty}(\psi^{ABC}) \geq \max \{ I(A\rangle B)_{\psi} , L(\psi) \},
\end{equation}
where $L(\psi):=\min \{I(AC\rangle B)_{\psi},I(A\rangle BC)_{\psi}\}$.
\end{theo}

{\bf Proof:} That $D_A^{\infty}(\psi^{ABC})$ is always greater than or equal to the coherent
information $I(A\rangle B)_{\psi}$ follows from the hashing inequality and the fact that Charlie's worst measurement is
no worse than throwing away his system and letting Alice and Bob perform a two-way distillation protocol without
outside help. Hence, it remains to show that $D_A^{\infty}(\psi^{ABC}) \geq \min \{I(AC\rangle B)_{\psi},I(A\rangle
BC)_{\psi}\}$.

Since $D^{\infty}_A(\psi^{ABC})$ is equal to the regularization of $D_A(\psi^{ABC})$, we only need to show the existence of a
measurement for Charlie for which the average distillable entanglement is asymptotically close to
$L(\psi)$. We prove this fact via a protocol which uses a random coding strategy. The state $\psi^{ABC}$ and its purifying system $R$ can be regarded as:
\begin{romanlist}
\item a tripartite system composed of $C,AB$, and $R$.
\item a tripartite system composed of $C,AR$, and $B$.
\end{romanlist}
 Let's consider $n$ copies of $\psi^{ABC}$, and furthermore, let's assume that $S(AB)_{\psi}$ (resp. $S(AR)_{\psi}$) and $S(R)_{\psi}$ (resp. $S(B)_{\psi}$) are different. This can be enforced by using only a sub-linear amount of entanglement shared between chosen parties in the limit of large $n$. After Schumacher compressing his share of the state $\psi_{C}^{\otimes n}$, Charlie performs a random measurement of his system $\tilde{C}$. Let $J$ be the random variable associated with the measurement outcome and let $\psi^{A^nB^nR^n}_J$ be the state of the systems $A^n, B^n$ and $R^n$ after Charlie's measurement. By Lemma \ref{lem:unionbound} and the Fannes inequality, there exists a measurement of Charlie's system which will produce a state $\psi^{A^nB^nR^n}_J$ satisfying, with arbitrarily high probability:
\begin{align} \label{eq:property}
  S(A^nB^n)_{\psi_J}=S(R^n)_{\psi_J} &= n(\min\{S(AB)_{\psi},S(R)_{\psi}\}
\pm \delta) \notag \\
  S(A^nB^n)_{\psi_J}=S(B^n)_{\psi_J} &= n(\min\{S(AR)_{\psi},S(B)_{\psi}\} \pm \delta),
\end{align}
where $\delta$ can be made arbitrarily small by choosing $n$ large enough.

Applying the hashing inequality to such a state will give:
\begin{align}
D(\psi^{A^nB^n}_J) & \geq S(B^n)_{\psi_J} - S(A^nB^n)_{\psi_J} \notag \\
&=  n(\min \{S(B)_{\psi},S(AR)_{\psi}\} \pm \delta) -
n(\min\{S(AB)_{\psi},S(R)_{\psi}\} \pm \delta) \notag \\
&\geq n (\min \{S(B)_{\psi},S(AR)_{\psi}\} - S(R)_{\psi} - 2\delta) \notag \\
& = n (\min \{ I(AC \rangle B)_{\psi},I(A \rangle BC)_{\psi} \} - 2\delta).
\end{align}
For each outcome $j$, define $X_j$ to be the variable taking the value zero if $\psi_j^{A^nB^nR^n}$ satisfies Eq.~(\ref{eq:property}), or one otherwise.
The average two-way distillable entanglement for this measurement will be at least
 \begin{align}
    \sum_j p_j D(\psi^{A^nB^n}_j) &= \sum_{X_j=0}p_j D(\psi^{A^nB^n}_j) + \sum_{X_j=1} p_j D(\psi^{A^nB^n}_j) \notag \\
    &\geq P(X_J=0)n (\min \{ I(AC \rangle B)_{\psi},I(A \rangle BC)_{\psi} \} - 2\delta) + \sum_{X_j=1} p_j D(\psi^{A^nB^n}_j) \notag \\
    &\geq (1-\alpha)n  \left[\min \{I(AC \rangle B)_{\psi},I(A \rangle
    BC)_ {\psi} \} - 2\delta\right],
 \end{align}
where $\alpha$ can be made arbitrarily small by taking sufficiently large values of $n$.  Finally, we have
 \begin{align}
   \frac{1}{n} D^A((\psi^{ABC})^{\otimes n}) &\geq \sum_j p_j D(\psi^{A^nB^n}_j) \notag \\
 &\geq (1-\alpha)\left[\min \{I(AC \rangle B)_{\psi},I(A \rangle BC)_ {\psi} \} -2\delta\right].
 \end{align}
Since $\alpha$ and $\delta$ can be chosen to be arbitrarily small, we are done. \square\,

\begin{coro} \label{cor:lowerbound}
Let $\psi^{ABC}$ be an arbitrary tripartite state shared by two recipients (Alice and Bob) and a helper (Charlie).
Then the asymptotic entanglement of assistance is bounded below as follows:
\begin{equation}\label{eq:L1}
D_A^{\infty}(\psi^{ABC}) \geq \lim_{n \rightarrow \infty}\frac{1}{n} \sup_{{\cal I}} \sum_i p_i L(\sigma^{A^nB^n\bar{C}}_i),
\end{equation}
where the supremum is over all instruments ${\cal I} := \{{\cal E}_i\}$ performed by Charlie, with $\sigma^{A^nB^n\bar{C}}_i = \frac{1}{p_i}(\mathrm{id}^{A^nB^n} \otimes {\cal E}_i)(\psi_{ABC}^{\otimes n})$ and $p_i = \Tr [ {\cal E}_i \psi_C^{\otimes n} ]$.
\end{coro}
{\bf Proof:}
First, to see that the maximization of Eq.~(\ref{eq:L}) can be removed, consider an instrument ${\cal J}$ which traces out the $C$ system: $\sigma^{AB} = \Tr_C \psi^{ABC}$. Then, both coherent information quantities in $L(\sigma)$ reduces to the coherent information $I(A\rangle B)_{\psi}$. Achievability of the rate $\sum_i p_i L(\sigma_i^{ABC})$, for any instrument ${\cal I}$ performed on $n$ copies of $\psi^{ABC}$, follows by considering a blocking strategy. \square\,

Let's look into some of the peculiarities of the previous results.
First, observe that the right hand side of Eq.~(\ref{eq:L}) is bounded from
above by the coherent information $I(A\rangle BC)_{\psi}$. This
follows from the definition of $L(\psi)$, and the strong
subadditivity of the von Neumann entropy, expressed in terms of coherent information quantities as: \begin{equation} I(A\rangle
BC)_{\psi} \geq I(A\rangle B)_{\psi}.\end{equation} When the lower bound of Eq.~(\ref{eq:L}) is equal to
$I(A\rangle BC)_{\psi}$, we have $I(A\rangle BC)_{\psi} \leq
I(AC\rangle B)_{\psi}$, which implies by further calculation that
$I(C\rangle B)_{\psi} \geq 0$. Suppose that $I(C \rangle B)_{\psi} > 0$ and consider $n$ copies of the purified state $(\psi^{ABCR})^{\otimes n}$, written
as $\psi^{C^nB^nR^n_1}$ where $R_1:=AR$ is the relative reference for the helper $C$. State merging~\cite{merge, SW-nature} tells us that a random measurement on
the typical subspace $\tilde{C}$, as described in our protocol, will decouple the system from its relative reference $R^n_1$, allowing recovery of $C^n$ by Bob up to arbitrarily high fidelity. Our assisted distillation protocol can be improved for this case by recovering the $C^n$ system at Bob's location
before engaging into a two-way distillation protocol, which will now act on the state
$(\psi^{AB\tilde{B}})^{\otimes n}$, where $\tilde{B}$ is an ancilla of the same dimension as the $C$ system. Since the distillable entanglement
across the cut $A$ vs $BC$ cannot increase by local operations and classical communication, the previous strategy is in fact optimal. A
small amount of initial entanglement between $C^n$ and $B^n$
may be needed if $I(C\rangle B)_{\psi}=0$ (see~\cite{merge}).

The previous analysis may lead us to believe that when the lower bound of Eq.~(\ref{eq:L}) is equal to $I(AC\rangle B)_{\psi}$, a similar strategy of
transferring the system $C$ to Alice could be
applied. However, the following counterexample will show that this
is not always true. Let
\begin{align}\label{eq:stateexample}
\ket{\psi}^{BC_2R}
    &= \frac{1}{\sqrt{2}} \ket{000}^{BC_2R}
            +\frac{1}{2}\ket{110}^{BC_2R}+\frac{1}{2}\ket{111}^{BC_2R} \quad \mbox{and} \notag \\
\ket{\psi}^{AC_1}
    &=  \frac{1}{2}\ket{00}^{AC_1} + \sqrt{\smfrac{3}{4}}\ket{11}^{AC_1}.
\end{align}
Alice and Bob are to perform assisted distillation on $n$ copies
of $\psi^{ABC_1C_2} = \psi^{AC_1} \otimes \psi^{BC_2}$ with the
help of a single Charlie holding both the $C_1$ and $C_2$ systems.
Such a situation could arise in practice if Alice had a
high-quality quantum channel to Charlie but the Charlie's channel
to Bob were noisy. The system $R$ would represent the environment of the
noisy channel.

In this case, $L(\psi)$ is equal to $I(AC\rangle B)_\psi$, which
is easily calculated to be approximately 0.40, since $I(A\rangle
BC)_\psi \approx 0.81$ and $I(A\rangle B)_\psi$ is negative.
For this example, the achievable rate of our random coding
protocol is therefore at least the rate that could have been
obtained by a strategy of first transferring the state of the $C$
system to Alice, followed by entanglement distillation between
Alice and Bob at the hashing bound rate. However, the coherent
information $I(C\rangle A)_{\psi}$ is negative for the state
$\psi^{ABC_1C_2R}$. By the optimality of state merging, the state
transfer from Charlie to Alice cannot be accomplished without the
injection of additional entanglement between them. Therefore, the
protocol achieves the rate $I(AC\rangle B)_\psi$ \emph{without}
performing the Charlie to Alice state transfer.

This example also illustrates a general relationship between
hierarchical distillation strategies and the random measurement
strategy proposed in this chapter. A hierarchical strategy for a
state $\psi^{ABC_1C_2} = \psi^{AC_1} \otimes \psi^{C_2B}$ would
consist of first distilling entanglement between $A$ and $C_1$ as
well as between $C_2$ and $B$, followed by entanglement swapping
to establish ebits between Alice and Bob. If the first level
distillations are performed at the hashing rate, then this
strategy will establish $\min [ I(A\rangle C_1)_\psi, I(C_2
\rangle B)_\psi ]$ ebits between Alice and Bob per copy of the input state. On the other hand,
the random measurement strategy will establish at least $L(\psi)$,
which in the case of the example is the minimum of
\begin{align}
I(AC \rangle B)_\psi
    &= I(C_2 \rangle B)_\psi - S(AC_1)_\psi  = I(C_2 \rangle B)_\psi \mbox{  and} \notag \\
I(A\rangle BC)_\psi
    &= I(A\rangle C_1)_\psi,
\end{align}
yielding exactly the same rate as the hierarchical strategy. (The
first line uses the fact that $\psi^{AC_1}$ is pure.) So, for the
random measurement strategy to beat the hierarchical strategy, it
is necessary that the state not factor into the form $\psi^{AC_1}
\otimes \psi^{C_2B}$. As an example, consider modifying the state of Eq.~(\ref{eq:stateexample}) by applying a controlled-NOT operation (CNOT) between
the systems $C_1$ and $C_2$ held by Charlie:
\begin{equation}
CNOT = \left( \begin{array}{cccc}
1 & 0 & 0 & 0 \\
0 & 1 & 0 & 0 \\
0 & 0 & 0 & 1 \\
0 & 0 & 1 & 0 \\
 \end{array} \right)
 \end{equation}
A CNOT operation can be used to model phase dampening effects between an input state (i.e the $C_2$ system) and its environment (i.e the $C_1$ system).
If the control qubit is the system $C_1$ and the target qubit is $C_2$, the previous state transforms to:
\begin{equation}
\ket{\phi}^{C_1C_2ABR}
    :=\frac{\ket{00}^{AC_1}\ket{\psi}}{2}^{BC_2R} + \sqrt{\smfrac{3}{4}}\ket{11}^{AC_1} \bigg ( \frac{\ket{010}}{\sqrt{2}}^{BC_2R}
            +\frac{\ket{100}}{2}^{BC_2R}+\frac{\ket{101}}{2}^{BC_2R} \bigg ).
\end{equation}
The reduced state $\phi^{C_1}$ of the $C_1$ system is equal to:
\begin{equation}
\phi^{C_1}:=\frac{1}{4}\braket{0}^{C_1} + \frac{3}{4}\braket{1}^{C_1},
\end{equation}
and the reduced state $\phi^{AC_1}$ of the system $AC_1$ is given by
\begin{equation}
\phi^{AC_1}:=\frac{1}{4}\braket{00}^{AC_1} + \frac{3}{4}\braket{11}^{AC_1}.
\end{equation}
Hence, the coherent information $I(A \rangle C_1)_{\phi}$ is zero, yielding a null rate for the hierarchical strategy. On the other hand, the quantities $I(AC \rangle B)_{\phi}$ and $I(A \rangle BC)_{\phi}$ are equal to the coherent informations $I(AC \rangle B)_{\psi}$ and $I(A \rangle BC)_{\psi}$. Thus, the random measurement strategy is unaffected by a CNOT ``error'' on Charlie's systems, as opposed to the hierarchical strategy, which fails to recover from this error.

Finally, it is easy to determine conditions under which
the random measurement strategy for assisted entanglement distillation will yield a
higher rate than the hashing bound between Alice and Bob. As the
next result shows, a state $\psi^{ABC}$ is a good
candidate for the random measurement strategy if it does not
saturate the strong subadditivity inequality of the von Neumann
entropy, and if the $C$ system can be redistributed to Alice and
Bob provided they are allowed to perform joint operations on their
systems.

\begin{prop}[Beating the Hashing Inequality]
\label{prop:abovehashing} For any state $\psi^{ABC}$, the value of
$L(\psi)$ is positive and strictly greater than the coherent
information $I(A \rangle B)_{\psi}$ if \begin{equation} I(C
\rangle AB)_{\psi}
> 0 \mbox{ and } S(A|BC)_{\psi} < S(A|B)_{\psi}.\end{equation}
\end{prop}
{\bf Proof:}
  The inequality $S(A|BC)_{\psi} < S(A|B)_{\psi}$ can be rewritten
  as \begin{equation}
    I(A\rangle BC)_{\psi} > I(A\rangle B)_{\psi},
    \end{equation} and the condition $I(C \rangle AB)_{\psi} >
  0$ as \begin{equation}
    S(AB)_{\psi} > S(ABC)_{\psi}.
    \end{equation} By
  negating and adding $S(B)_{\psi}$ on both sides of the previous inequality, we
  get back
  \begin{equation}
      I(A\rangle B)_{\psi} := S(B)_{\psi} - S(AB)_{\psi} < S(B)_{\psi} - S(ABC)_{\psi} =: I(AC \rangle B)_{\psi}.
  \end{equation}
\square\,

\section{Multipartite Entanglement of Assistance} \label{sec:multi}
\begin{figure} [t]
\centerline{\epsfig{file=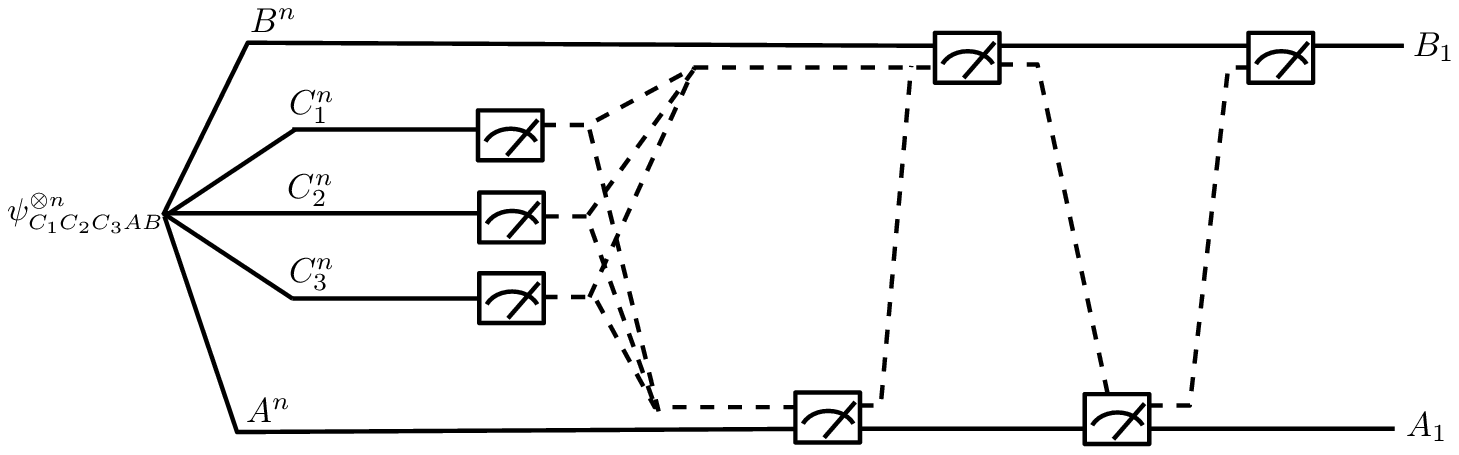, width=15.2cm}} 
\vspace*{13pt}
\fcaption{\label{fig:many}Quantum circuit representing a broadcast assisted entanglement distillation protocol involving three helpers. The three helpers perform their measurements, sending copies of the classical outcomes to Alice and Bob. Alice and Bob then implement an LOCC operation, based on that outcome.}
\end{figure}

\noindent
In this section, we look at the optimal rate achievable when many spatially separated parties are assisting
Alice and Bob in distilling entanglement. First, we extend the one-shot entanglement of assistance $D_A$ (Definition \ref{def:A1D}) to arbitrary multipartite states $\psi^{C_1C_2\ldots C_mAB}$, henceforth written simply as $\psi^{C_MAB}$. The type of protocols involved is depicted in Figure~\ref{fig:many}.
\begin{defi}
For a general multipartite state $\psi^{C_MAB}$, consider POVMs $E_1, \ldots, E_m$ performed by
$\{C_1,C_2,\ldots,C_{m}\}$ respectively which lead to a (possibly mixed) bipartite state $\psi^{AB}_{k_1k_2\ldots k_m}$ for POVM outcomes $\overline{k}:=k_1k_2\ldots k_m$. We define the multipartite entanglement of assistance as
\begin{equation}
D_A(\psi^{C_MAB}) := \sup \sum_{\overline{k}} p_{\overline{k}} D(\psi^{AB}_{\overline{k}}),
\end{equation}
where the supremum is taken over the above measurements. The asymptotic multipartite entanglement of
assistance $D^{\infty}_A(\psi^{C_MAB})$ is obtained by regularization of the above quantity
$D_A^{\infty}(\psi^{C_MAB})=\lim_{n\rightarrow\infty} \frac{1}{n} D_A(\psi^{\otimes n})$.
\end{defi}

For a pure state $\psi^{C_MAB}$, it is immediate that the supremum in the preceding definition is attained for POVMs of rank one, leading to an ensemble of pure states $\{q_{\overline{k}}, \psi^{AB}_{\overline{k}}\}$. And so,  $D^{\infty}_A(\psi^{C_MAB})$ reduces to the asymptotic multipartite entanglement of assistance~\cite{merge} for pure states:
\begin{equation}
\label{eq:mincutEofA}
D_A^{\infty}(\psi^{C_MAB}) = \min_{{\cal T}} \{ S(A{\cal T})_{\psi}\},
\end{equation}
where the minimum is taken over all bipartite cuts $\cal{T}$. (A bipartite cut consists of a partition of the helpers ${C_1,\ldots, C_{m}}$
into a set $\cal{T}$ and its complement $\overline{\cal T }=\{C_1,\ldots, C_{m}\} \backslash \cal{T}$.)

\begin{prop}\label{prop:upper}
Let $\psi^{C_MAB}$ be an arbitrary multipartite state. The quantity $D^{\infty}_A(\psi^{C_MAB})$ is bounded
from above by the following quantity:
\begin{equation}
  D^{\infty}_A(\psi^{C_MAB}) \leq \min_{{\cal T}}  D(\psi^{A{\cal T}|B\overline{{\cal T}}}),
\end{equation}
where the minimum is over all bipartite cuts $\cal{T}$ and $\psi^{A{\cal T}|B\overline{{\cal T}}}$ is a bipartite state with Alice holding the systems $A{\cal T}$ and Bob holding the systems $B\overline{\cal T}$.
\end{prop}
{\bf Proof:} Consider any cut ${\cal T}$ of the helpers $\{C_1,C_2,\ldots, C_{m}\}$ and suppose Alice (resp. Bob) is allowed to perform joint operations on the systems $A{\cal T}$ (resp. $B{\overline{\cal T}}$). Any protocol achieving $D^{\infty}_A(\psi^{C_MAB})$ consists of: 1) POVMs on the helpers followed by a transmission of the outcomes to Alice and Bob 2) local operations and classical communication between the systems $A$ and $B$. This kind of protocol is contained in protocols allowing local operations on the systems $A{\cal T}$ and $B{\overline{{\cal T}}}$  and classical communication between the cut $A{\cal T}$ vs $B{\cal \overline{T}}$. Since the distillable entanglement across the cut $A{\cal T}$ vs $B{\cal {\overline{T}}}$ cannot increase under local operations and classical communication, the optimal achievable rate for these protocols is given by $D(\psi^{A{\cal T}|B{\overline{{\cal T}}}})$. Since this holds for any cut ${\cal T}$ of the helpers, we are done. \square\,

\begin{defi}
\label{minhashcut} For an arbitrary multipartite state $\psi^{C_MAB}$, we
define the minimum cut coherent information as:
\begin{equation} I^c_{min}(\psi,A:B) := \min_{{\cal T}} I(A {\cal T} \rangle B\overline{{\cal T}})_{\psi},
\end{equation} where the minimization is over all bipartite cuts ${\cal T} \subseteq \{C_1,C_2,\ldots, C_{m}\}$.
\end{defi}

\begin{theo}\label{thm:gen}
Let $\psi^{C_MAB}$ be an arbitrary multipartite state. The asymptotic
multipartite entanglement of assistance $D_A^{\infty}(\psi^{C_MAB})$ is bounded below by:
\begin{equation}
\label{eq:lowerbound} D_A^{\infty}(\psi^{C_MAB}) \geq \max\{I(A\rangle B)_{\psi},I^c_{min}(\psi, A:B)\}.
\end{equation}
\end{theo}
Before giving a proof of Theorem \ref{thm:gen}, we need the following lemma, which states that the minimum cut coherent information of the original state is preserved, up to a vanishingly small perturbation, after an helper has finished performing a random measurement on his system. The arguments needed for  demonstrating this lemma are similar to those used in~\cite{merge} to prove Eq.~(\ref{eq:mincutEofA}).
\begin{lem}\label{lem:mincut}
Given $n$ copies of a state $\psi^{C_MAB}$, let $C_{m}$ perform a random measurement on his typical subspace $\tilde{C}_{m}$ as in Lemma~\ref{lem:unionbound}. For any $\delta > 0$ and large enough $n$, there exists a measurement performed by the helper $C_m$ such that, with arbitrarily high probability, the outcome state $\psi^{C_1^n\ldots C_{m-1}^nA^nB^n}_J$ satisfies the following
inequality:
\begin{equation}
I^c_{min}(\psi_J,A^n:B^n) \geq n (I^c_{min}(\psi,A:B) - \delta),
\end{equation}
where $J$ is the random variable associated with the measurement outcome.
\end{lem}
{\bf Proof:} The minimum cut coherent information $I^c_{min}(\psi,A:B)$ of the state $\psi^{C_MAB}$ can be rewritten as
   \begin{equation}
      I^c_{min}(\psi, A:B) = \min_{{\cal T} \subseteq \{C_1,C_2,\ldots, C_{m}\}} \bigl \{ S(B{\cal T})_{\psi} \bigr \} - S(R)_{\psi},
   \end{equation}
where $R$ is the purifying system for the state $\psi^{C_MAB}$.

Let $\cal T$ be a bipartite cut of the helpers $\{C_1,C_2,\ldots, C_{m}\}$ such that $C_{m} \notin
{\cal T}$. We define its relative complement as ${\cal T'}=\{C_1,\ldots,C_{m-1}\} \backslash \cal T$.
For any such cut $\cal T$, the state $\psi^{C_MAB}$ and its purifying system $R$ can be regarded as a tripartite system composed of $C_{m}$, $AR{\cal T}$ and $B{\cal T'}$. Assuming $S(AR{\cal T})_{\psi}$ and $S(B\cal T')_{\psi}$ to be distinct, the helper $C_{m}$ performs a random measurement on his typical subspace $\tilde{C}_m$. By Proposition \ref{thm:random} (see Appendix A) and the Fannes inequality, there exists a measurement for Charlie's system for which the outcome state $\psi^{C_1^n\ldots C_{m-1}^nA^nB^n}_J$ satisfies, with arbitrarily high probability:
\begin{equation}
 \label{eq:entropy1}
\min\{S(AR{\cal T})_{\psi},S(B{\cal T'})_{\psi}\} - \delta' \leq \frac{1}{n} S(B^n{\cal T'}^n)_{\psi_J} \leq \min\{S(AR{\cal T})_{\psi},S(B{\cal T'})_{\psi}\} + \delta',
\end{equation} where $\delta'$ can be made arbitrarily small by taking sufficiently large values for $n$. Hence, the reduced state entropies stay distinct by taking a sufficiently small value of $\delta'$. Since $I^c_{min}(\psi_J,A^n:B^n)$ can be re-expressed as
\begin{equation}
\label{mincut} I^c_{min}(\psi_J,A^n:B^n) = \min_{{\cal T}\subseteq \{C_1,C_2,\ldots, C_{m-1}\}} \{S(B^n{\cal T'}^n)_{\psi_J} \} - S(R^n)_{\psi_{J}},
\end{equation}
we can substitute the lower bound for $S(B^n{\cal T'}^n)_{\psi_J}$ into Eq.~(\ref{mincut}) and obtain
\begin{align}
I^c_{min}(\psi_J,A^n:B^n) & \geq n \min_{\cal T}  ( \min\{S(AR{\cal
T})_{\psi},S(B{\cal T'})_{\psi}    \} - \delta' ) - S(R^n)_{\psi_{J}}\notag \\
 & = n (\min_{\cal T} \{ S(B{\cal T'}C_{m})_{\psi},S(B{\cal T'})_{\psi} \} - \delta') -
S(R^n)_{\psi_J} \notag \\
 & = n (\min_{\cal T} \{ S(B{\cal T})_{\psi}\} - \delta')  -
S(R^n)_{\psi_J}. \label{psijeq}
\end{align}

To finish the proof, the last fact we need concerns the entropy of the purifying system $R$. If we consider the purified state
$\psi^{C_MABR}$ as a tripartite system composed of $C_{m}$, $R$ and $ABC_1,\ldots,C_{m-1}$, we
can apply Proposition \ref{thm:random} and obtain, w.h.p:
\begin{align}
 \label{eq:entropyreference}
S(R^n)_{\psi_J} & =  n
(\min\{S(R)_{\psi},S(C_1,\ldots,C_{m-1}AB)_{\psi}\} \pm \delta'' )\notag \\
& \leq  n(S(R)_{\psi} + \delta''),
\end{align}
where $\delta''$ can be made arbitrarily small. This tells us that for large values of $n$, the entropy of the purifying system will not significantly increase as a result of the helper $C_m$ performing a measurement on his typical subspace $\tilde{C}_{m}$. Note that, as in Theorem \ref{thm:lowerbound}, we can use the union bound and Markov's inequality (see Lemma \ref{lem:unionbound}) to show the existence of a measurement on $\tilde{C}_{m}$ which produces states such that, w.h.p, Eqs.(\ref{eq:entropy1}) and (\ref{eq:entropyreference}) are both satisfied.
Combining the last equation with Eq.~(\ref{psijeq}) and choosing values
for $\delta',\delta''$ small enough that $\delta'+\delta'' < \delta$, we get the desired result. \square\,

\smallskip
\noindent {\bf Proof of Theorem~\ref{thm:gen}:} The right hand side of Eq.~(\ref{eq:lowerbound}) is just the coherent information when $m=0$, and is equal to $\max\{I(A \rangle B)_{\psi}, L(\psi)\}$ for $m=1$. Eq.~(\ref{eq:lowerbound}) holds for these base cases by the hashing inequality and Theorem \ref{thm:lowerbound}. So, from here on, assume $m \geq 2$. Moreover, that $D^{\infty}_A(\psi^{C_MAB})$ is at least $I(A\rangle B)_{\psi}$ follows again from the hashing inequality. Hence, we can focus on proving that $D^{\infty}_A(\psi^{C_MAB})$ is bounded below by the minimum cut coherent information.

By Lemma \ref{lem:mincut}, there exists a measurement $E_m$ for the helper $C_m$ which produces an outcome state $\psi^{C_1^n\ldots C_{m-1}^nA^nB^n}_{J}$ satisfying w.h.p. the following inequality:
\begin{equation}
 \begin{split}
I^c_{min}(\psi_J,A^n:B^n) \geq n (I^c_{min}(\psi,A:B)-\delta), \\
 \end{split}
\end{equation}
for an arbitrary small $\delta$ and sufficiently large $n$. If we have at our disposal $n^{m}$ copies of the state $\psi^{C_MAB}$ and perform the measurement $E_m$ for each block of $n$ copies of the state, we expect to obtain approximately $n^{m-1}p_j$ copies of the state $\psi_j$, with $p_j$ being the probability of obtaining the state $\psi_j$ after the measurement $E_m$ is performed by $C_{m}$.

For each block consisting of many copies of the state $\psi_j$, we repeat the previous procedure in a recursive manner. We continue this process until all $m$ helpers have performed measurements on their systems. In the end, Alice and Bob will obtain a number of bipartite states $\psi^{AB}_{J_1J_2\ldots J_{m}}$ each satisfying w.h.p.
\begin{equation}
I^c_{min}(\psi^{AB}_{J_1J_2\ldots J_{m}},A^{n^{m}}:B^{n^{m}}) \geq n^{m} (I^c_{min}(\psi,A:B)-\delta'), \\
\end{equation}
where $\delta'$ can be made arbitrarily small. Observe that the term on the left hand side of the inequality is the coherent information $I(A\rangle B)_{\psi^{AB}_{J_1J_2\ldots J_{m}}}$, which is bounded above by the distillable entanglement $D(\psi^{AB}_{J_1J_2\ldots J_{m}})$. Since $D_A(\psi^{n^{m}})$ is a supremum over all LOCC measurements performed by the helpers, we have
 \begin{align}
  \frac{1}{n^{m}} D_A(\psi^{\otimes (n^{m})}) &\geq \frac{1}{n^{m}} \sum_{j_1j_2\ldots j_m} p_{j_1j_2\ldots j_m} D(\psi^{AB}_{j_1j_2\ldots j_{m}})\notag \\
  &\geq \frac{1}{n^{m}}\sum_{j_1j_2\ldots j_m} p_{j_1j_2\ldots j_m} I(A\rangle B)_{\psi^{AB}_{j_1j_2\ldots j_{m}}} \notag \\
  &\geq (1-\epsilon)(I^c_{min}(\psi,A:B)-\delta'),
\end{align}
where $\epsilon$ and $\delta'$ can be both be made arbitrarily small by the arguments of the previous paragraphs. This concludes the proof. \square\,

\begin{figure} [t]
\centerline{\epsfig{file=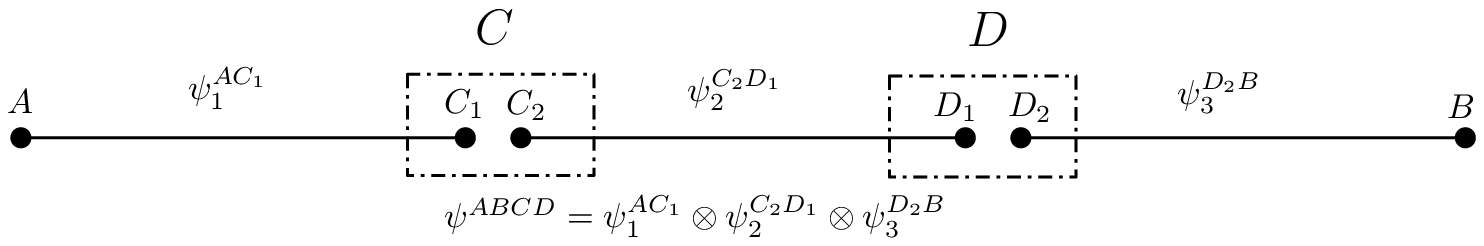, width=14.2cm}} 
\vspace*{13pt}
\fcaption{\label{fig:chain}A 1-dimensional chain with two repeater stations
separating the two recipients $A$ and $B$.}
\end{figure}

Before closing this section, let us say a few words on assisted
distillation when the two recipients are separated by a
one-dimensional chain of repeater nodes, as depicted in figure
\ref{fig:chain}. Applying a hierarchical distillation strategy on
$\psi^{ABCD}$ will produce ebits at  the rate
\begin{equation}
 R(\psi):=\min\{I(A\rangle C_1)_{\psi_1},I(C_2 \rangle
D_1)_{\psi_2}, I(D_2 \rangle B)_{\psi_3}\}.
\end{equation}

If we consider the cut ${\cal T}_1:=\{C\}$ of the helpers $C$ and
$D$, the coherent information $I(A{\cal T}_1\rangle
B{\overline{\cal T}_1})_{\psi}$ can be simplified to
\begin{equation} I(A{\cal T}_1\rangle B{\overline{\cal T}_1})_{\psi}
= I(C_2 \rangle D_1)_{\psi_2} - S(AC_1)_{\psi_1} \leq I(C_2
\rangle D_1)_{\psi_2},
\end{equation} and similarly for the cuts
${\cal T}_2=\{CD\}$ and ${\cal T}_3=\emptyset$, we have
\begin{align}
I(A{\cal T}_2\rangle B{\overline{\cal T}_2})_{\psi} &= I(D_2
\rangle B)_{\psi_3} - S(AC_1)_{\psi_1} - S(C_2D_1)_{\psi_2} \leq
I(D_2\rangle B)_{\psi_3}, \notag \\
I(A{\cal T}_3\rangle B{\overline{\cal T}_3})_{\psi} &= I(A \rangle C_1)_{\psi_1}.
\end{align} Thus, the minimum cut coherent information
$I^c_{min}(\psi^{ABCD},A:B)$ is not greater than $R(\psi)$, and so
a hierarchical strategy might be better suited for the case of a
chain state than a random measurement strategy, provided the information stored in the repeaters is not subject to errors (see the example of the previous section). It is easy to
generalize the previous arguments to a chain of arbitrary length
(i.e $m \geq 3$), and to other network configurations.

\section{Discussion} \label{sec:discussion}
\noindent
We generalized the entanglement of assistance problem by allowing the parties to share a multipartite mixed state. For the case of three parties holding a mixed tripartite state, the optimal assisted distillation rate was proven to be equal to the regularization of the one-shot entanglement of assistance, a quantity which maximizes the average distillable entanglement over all measurements performed by the helper. Two upper bounds for this quantity were established and examples of classes of states attaining them were given. Additionally, the one-shot entanglement of assistance was proven to be a convex quantity for pure ensembles.

We also presented new protocols for assisted entanglement distillation, based on a random coding strategy, which are proven to distill entanglement at a rate no less than the minimum cut coherent information, defined as the minimum coherent information over all possible bipartite cuts of the helpers. For states not saturating strong subadditivity, and recoverable by Alice and Bob if they can implement joint operations, we proved that our random coding strategy achieves rates surpassing the hashing inequality. Moreover, the rates formally resemble those achievable if the helper system were merged to either Alice or Bob even when such merging is impossible. Finally, we compared our protocol to a hierarchical strategy in the context of quantum repeaters. We identified a major weakness of the hierarchical strategy by analyzing the effect of a CNOT error on the rates achievable for such strategy. We found that the rate, which can be as good as the rate of our random measurement, becomes null when such error occurs at the repeater node. On the other hand, our protocol is completely fault tolerant and yields the same rate even if this error goes undetected by the helper holding the systems at the repeater node.

Our proposed protocol for assisted distillation of a general multipartite state, involved a measurement on a long block of states, and then a measurement on blocks of these blocks, and so on. It seems likely that a strategy where all the helpers measure in a random basis of their respective typical subspaces and broadcast the results to Alice and Bob will still yield a rate attaining at least the minimum cut coherent information. For pure states $\psi^{ABC_1C_2}$, we show in~\cite{Dutil1} that such a strategy is indeed possible by introducing a state merging protocol which acts on a pure multipartite state. However, for a general multipartite state, it is still unknown if such a strategy would work.

We could extend our assisted entanglement scenario in several ways. For instance, we could consider other forms of pure entanglement such as Greenberger-Horne-Zeilinger (GHZ) states and look at the optimal achievable rates under LOCC operations. Another interesting question is to analyze whether general LOCC operations between the parties give more power to the helpers. In the pure multipartite case, we saw that such strategy is not required to achieve optimal assisted rates. For multipartite mixed state, we should expect a difference in achievable rates when allowing more communication freedom to the helpers. We just have to consider bipartite distillation protocols to see this: the hashing protocol is impossible without communication between the parties. Finally another potential line of research is to analyze our assisted protocol using smooth min and max entropies. Recent work by Buscemi and Datta~\cite{Datta2} analyzed the one-way distillable entanglement of a bipartite mixed state $\psi^{AB}$ in the one-shot regime using one-shot entropic quantities similar to the quantum min- and max-entropy of~\cite{Renner02}. The entanglement of assistance for pure states $\psi^{ABC}$ was also analyzed under this framework~\cite{Datta}, and it is another natural progression of our work to analyze the entanglement of assistance using the quantum min- and max-entropy formalism.

\nonumsection{Acknowledgements}
\noindent
The authors would like to thank Mark Wilde for his helpful comments.
This research was supported by the Canada Research Chairs
program, CIFAR, FQRNT, INTRIQ, MITACS, NSERC, ONR grant
No.~N000140811249 and QuantumWorks.

\nonumsection{References}
\noindent
\bibliographystyle{unsrt}
\bibliography{EofABib}{}

\appendix
\noindent
 \subsection{Miscellaneous facts}
 The trace distance between two density operators $\rho$ and $\sigma$ is defined to be $D(\rho,\sigma) = \frac{1}{2}\|\rho-\sigma\|_1$. The trace distance and fidelity are related as follows:
   \begin{lem}\cite{FuchsVandegraaf:fidelity}\label{eq:purified}
   For states $\rho$ and $\sigma$, the trace distance is bounded by
         \begin{equation} 1- F(\rho, \sigma) \leq D(\rho,\sigma) \leq \sqrt{1-F^2(\rho,\sigma)}. \end{equation}
   \end{lem}

\begin{lem}[Fannes Inequality]\label{lem:Fannes}
     Let $\rho^A$ and $\sigma^A$ be states on a $d$-dimensional Hilbert space $A$. Let $\epsilon > 0$ be such that $\|\rho^A - \sigma^A\|_1 \leq \epsilon$. Then \begin{equation}
     \bigg |S(A)_{\rho} - S(A)_{\sigma} \bigg | \leq \eta(\epsilon) \log{d},
     \end{equation}where $\eta(x) = x - x \log x$ for $x \leq \frac{1}{e}$. When $x > \frac{1}{e}$, we set $\eta(x)=x + \frac{\log(e)}{e}$.
\end{lem}
For two sub-normalized states $\rho$ and $\bar{\rho}$, we define the \textit{purified distance}~\cite{Renner01} between $\rho$ and $\bar{\rho}$ as
\begin{equation}
    P(\rho,\bar{\rho}) := \sqrt{1- \overline{F}(\rho,\bar{\rho})^2},
 \end{equation}
where $\overline{F}(\rho,\bar{\rho})$ is the generalized fidelity~\cite{Renner01} between $\rho$ and $\bar{\rho}$:
\begin{equation}
   \overline{F}(\rho,\bar{\rho}) := F(\rho, \bar{\rho}) + \sqrt{(1-\Tr \rho)(1- \Tr\bar{\rho})}.
\end{equation}

\begin{lem}\cite{Renner01}\label{lem:purdist}
For sub-normalized states $\rho$ and $\bar{\rho}$, the trace distance is related to the purified distance as follows
 \begin{equation}
    D(\rho, \bar{\rho}) \leq P(\rho, \bar{\rho}) \leq 2 \sqrt{D(\rho,\bar{\rho})}.\label{eq:purified}
 \end{equation}
\end{lem}
A proof of this fact follows directly from Lemma 6 of \cite{Renner01}. 

Let $\psi^A = \sum^{X}_{x=1} p_x \braket{x}^A$ and fix any $\delta > 0$. For a sequence $x^n$ of $n$ letters $x_1x_2x_3\ldots x_n$, where each letter is taken from an alphabet ${\cal X}$ of size $X$, let $N(x_i|x^n)$ be the number of times the letter $x_i$ appears in the sequence $x^n$. We define the $\delta-$\textit{typical subspace} $\tilde{A}^n_{\psi,\delta}$ for the density operator $\psi_A^{\otimes n}$ as
\begin{equation}
  \tilde{A}^n_{\psi,\delta} := \mathrm{span} \bigg \{ \ket{x^n} \bigg | x^n \in {\cal T}^n_{p,\delta} \bigg \},
\end{equation}
where 
\begin{equation}
   {\cal T}^n_{p,\delta} := \bigg \{ x^n : \forall x \in {\cal X}, \bigg | \frac{N(x_i|x^n)}{n} - p(x) \bigg | \leq \delta \bigg \}.
\end{equation}
The projector into the typical subspace $\tilde{A}^n_{\psi,\delta}$ is given by:
\begin{equation*}
  \Pi^n_{\psi,\delta} := \sum_{x^n \in {\cal T}^n_{p,\delta}} \braket{x^n}.
\end{equation*}
We abbreviate the $\delta-$typical subspace $\tilde{A}^n_{\psi,\delta}$ associated with the state $\psi^{\otimes n}_A$ as $\tilde{A}$ and the typical projector $\Pi^n_{\psi,\delta}$ as $\Pi_{\tilde{A}}$.

\begin{lem}[Markov's Inequality] \label{lem:Markov}
 If $X$ is a random variable with probability distribution $p(x)$ and expectation $E(X)$, then, for any positive number $a$, we have:
 \begin{equation}
 P(|X| \geq a) \leq \frac{E(|X|)}{a}.
 \end{equation}
  \end{lem}
\subsection{Convexity of $D_A$ for pure ensembles}
\begin{lem}
For a state $\psi^{ABC}=\sum_i p_i \psi^{ABC}_i$, where $\psi^{ABC}_i$ are pure states, let $F =\{F_x\}_{x=1}^X$ be a POVM of rank one operators on the system $C$. Then, we have
\begin{equation}
   \sum_x q_x D(\psi^{AB}_x) \leq \sum_{x,i} p_i \Tr[F_x\psi_i^{C}] D(\tilde{\psi}^{AB}_{i,x}) ,
\end{equation}
where $\psi^{AB}_x = \frac{1}{q_x} \Tr_C[ (F_x \otimes I^{AB})\psi^{ABC}]$, $q_x = \Tr[F_x \psi^C]$ and $\tilde{\psi}^{AB}_{i,x} = \frac{1}{\Tr[F_x \psi_i^C]} \Tr_C[ (F_x \otimes I^{AB}) \psi_i^{ABC}]$.
\end{lem}

{\bf Proof:} The state we get after applying $F_x$ on system $C$ is given by:
 \begin{align}\label{eq:pureD}
   \frac{1}{q_x} \Tr_{C} \biggl [ \biggl (F_x \otimes I^{AB} \biggr ) \sum_i p_i \psi_i^{ABC} \biggr ] &= \frac{1}{q_x} \sum_i p_i \Tr_C[(F_x \otimes I^{AB}) \psi_i^{ABC}] \notag\\
    &= \sum_i \frac{p_i \Tr_C[F_x \psi_i^C]}{q_x} \tilde{\psi}^{AB}_{i,x}.
 \end{align}
Since $q_x = \Tr_C[F_x \psi^{C}]$ and $\sum_i p_i \psi^C_i = \psi^C$, we have a well-defined ensemble of pure states on the right hand side of Eq.~(\ref{eq:pureD}). Since the distillable entanglement is bounded from above by the entanglement of formation, we get
 \begin{align}
   \sum_x q_x D( \frac{1}{q_x} \Tr_{C} \biggl [ \biggl (F_x \otimes I^{AB} \biggr ) \sum_i p_i \psi_i^{ABC} \biggr ]) &\leq \sum_x q_x \sum_i \frac{p_i \Tr_C[F_x \psi_i^C]}{q_x} S(A)_{\tilde{\psi}_{i,x}} \notag \\
   &= \sum_{x,i} p_i \Tr_C[F_x \psi_i^C] D(\tilde{\psi}^{AB}_{i,x}).
 \end{align}
\square\,

\begin{prop}[Convexity of $D_A$ for Pure Ensembles]\label{lemma:convex}
Let $\psi^{ABC}$ be an arbitrary tripartite state. Then, for any convex decomposition $\{p_i,
\psi_i^{ABC}\}$ of $\psi^{ABC}$ into pure states,
  \begin{equation}
    D_A(\psi^{ABC}) \leq \sum_i p_i D_A(\psi_i^{ABC}).
  \end{equation}
\end{prop}
{\bf Proof:} For any $\nu > 0$, there exists a POVM $E=\{E_x\}^X_{x=1}$ of rank one operators such that
\begin{equation}
  \sum_x q_x D(\psi^{AB}_x) \geq D_A(\psi^{ABC}) - \nu,
\end{equation}
where $\psi^{AB}_x = \frac{1}{q_x} \Tr_{C}[(E_x \otimes I^{AB}) \psi^{ABC}]$.
From the previous lemma, we have
 \begin{align}
      \sum_x q_x D(\psi^{AB}_x) &\leq \sum_{x,i} p_i \Tr[E_x\psi_i^{C}] D(\tilde{\psi}^{AB}_{i,x}) \notag \\
        &\leq \sum_i p_i \sum_x  \Tr[E_x\psi_i^{C}] D(\tilde{\psi}^{AB}_{i,x}) \notag \\
        &\leq \sum_i p_i D_A(\psi_i^{ABC}),
 \end{align}
where $\tilde{\psi}_{i,x} = \frac{1}{\Tr_C[E_x \psi_i^{C}]}\Tr_{C}[(E_x \otimes I^{AB})\psi^{ABC}_i]$ is the state obtained after performing the POVM $E$ on the state $\psi^{ABC}_i$. Since $\nu$ was arbitrarily chosen, we get back the statement of the proof. \square\,

\subsection{Proof of Lemma \ref{lem:unionbound}}
Before proving Lemma \ref{lem:unionbound}, we state the result of~\cite{merge} used to prove the formula, Eq.~(\ref{eq:mincutEofA}), for the entanglement of assistance of pure multipartite states.
\begin{prop}\cite{merge} \label{thm:random} Suppose we have $n$ copies of a tripartite pure state $\psi^{CBR}$, where $S(R)_{\psi} < S(B)_{\psi}$.
Let $\psi^{\tilde{C}\tilde{B}\tilde{R}}$  be the normalized state obtained by projecting
$C^n, B^n, R^n$ into their respective typical subspaces $\tilde{C},\tilde{B},\tilde{R}$. Charlie performs a projective measurement using an orthonormal basis $\{\ket{e_i}^{\tilde{C}}\}$ of $\tilde{C}$ chosen at random according to the Haar measure. Denote by $p_i$ the probability of obtaining outcome $i$. Then, for any $\epsilon > 0$, and large enough $n$, we have
\begin{equation}
\int_{\mathbb{U}(\tilde{C})}  \sum_i p_i \bigl \| \psi^{\tilde{R}}_i - \psi^{\tilde{R}} \bigr \|_1 dU \leq \epsilon,
\end{equation}
where $\psi^{\tilde{R}}_i$ is the state of the system $\tilde{R}$ upon
obtaining outcome $i$. The average is taken over the unitary group $\mathbb{U}(\tilde{C})$ using the Haar measure.
\end{prop}

{\bf Proof of Lemma \ref{lem:unionbound}:}
The proof of this statement is obtained by combining Proposition \ref{thm:random} with Markov's inequality and Boole's inequality (the union bound). For any $\xi_1 > 0$ and $\xi_2 > 0$, consider a projective measurement of Charlie with rank one projectors $U\braket{i}U^{\dag}$ and let $J$ be the measurement outcome. We want to bound the following probability from below:
\begin{equation}\label{eq:probR}
P_J:=P(\| \psi^{R^n}_J - (\psi^R)^{\otimes n} \|_1 < \xi_1 \bigcap \| \psi^{B^n}_J - (\psi^B)^{\otimes n}\|_1 < \xi_2)  \geq 1 - \alpha
\end{equation}
for any $\alpha > 0$. Applying the union bound and Markov's inequality to such probability, we have
\begin{align}\label{eq:probS}
P_J &\geq 1 - P(\| \psi^{R^n}_J - (\psi^R)^{\otimes n} \|_1 \geq \xi_1) - P(\| \psi^{B^n}_J - (\psi^B)^{\otimes n}\|_1 \geq \xi_2)  \notag \\
&\geq 1- \frac{\sum_j p_j \| \psi^{R^n}_j - (\psi^R)^{\otimes n} \|_1} {\xi_1} - \frac{\sum_j p_j \| \psi^{B^n}_j - (\psi^B)^{\otimes n} \|_1}{\xi_2}.
\end{align}
Taking the average over the unitary group $\mathbb{U}(\tilde{C})$ using the Haar measure, we get
\begin{equation}\label{eq:probS}
\int_{\mathbb{U}(\tilde{C})} P_J dU \geq 1- \frac{\int_{\mathbb{U}(\tilde{C})} \sum_j p_j \| \psi^{R^n}_j - (\psi^R)^{\otimes n} \|_1 dU} {\xi_1} - \frac{\int_{\mathbb{U}(\tilde{C})}\sum_j p_j \| \psi^{B^n}_j - (\psi^B)^{\otimes n} \|_1 dU}{\xi_2}.
\end{equation}
The averages are not quite of the desired form to apply Proposition \ref{thm:random} directly. Define the state
\begin{equation}
\ket{\Omega}^{\tilde{C}\tilde{A}\tilde{B}\tilde{R}} := (\Pi_{\tilde{A}} \otimes \Pi_{\tilde{B}} \otimes \Pi_{\tilde{C}} \otimes \Pi_{\tilde{R}}) \ket{\psi}^{\otimes n},
\end{equation}
and let $\ket{\Psi}^{\tilde{C}\tilde{A}\tilde{B}\tilde{R}}$ be the normalized version of $\ket{\Omega}^{\tilde{C}\tilde{A}\tilde{B}\tilde{R}}$. If Charlie were to perform his measurement on the state $\Psi$, the properties of typicality tell us that the trace norms $\|\psi^{R^n}_j - \Psi^{\tilde{R}}_j\|_1$  and $\|\psi^{B^n}_j - \Psi^{\tilde{B}}_j\|_1$ should be arbitrarily close. This is verified by using the bounds between the trace distance and the purified distance , Lemma~\ref{lem:purdist}, and the monotonicity of the purified distance under trace non-increasing quantum operations (see~\cite{Renner01} for a proof of this fact):
\begin{align}
 \|\psi^{R^n}_j -  \Psi^{\tilde{R}}_j\|_1 &\leq \|\psi_j^{\tilde{C}A^nB^nR^n} - \Psi_j^{\tilde{C}\tilde{A}\tilde{B}\tilde{R}}\|_1\notag \\
 &\leq 2P(\psi^{\tilde{C}A^nB^nR^n}_j, \Psi^{\tilde{C}\tilde{A}\tilde{B}\tilde{R}}_j) \notag \\
&\leq 2P(\psi^{\tilde{C}A^nB^nR^n}, \Psi^{\tilde{C}\tilde{A}\tilde{B}\tilde{R}}) \notag \\
&\leq \epsilon,
\end{align}
for any $\epsilon > 0$ by choosing sufficiently large values of $n$. The last line follows from typicality and the triangle inequality. A similar statement holds for the trace norm $\|\psi^{B^n}_j - \Psi^{\tilde{B}}_j\|_1$.
Applying the triangle inequality twice on each average of Eq.~(\ref{eq:probS}), we have
\begin{equation}\label{eq:probSD}
\int_{\mathbb{U}(\tilde{C})} P_J dU  \geq 1- f(\epsilon) - \frac{\int_{\mathbb{U}(\tilde{C})} \sum_j p_j \| \Psi^{\tilde{R}}_j - \Psi^{\tilde{R}} \|_1 dU}{\xi_1} - \frac{\int_{\mathbb{U}(\tilde{C})} \sum_j p_j \| \Psi^{\tilde{B}}_j - \Psi^{\tilde{B}}\|_1 dU}{\xi_2},
\end{equation}
where $f(\epsilon)$ is a function of various trace norms which vanish, by typicality, for sufficiently large values of $n$. Applying Proposition \ref{thm:random} on the averages of Eq.~(\ref{eq:probSD}), we can make the right hand side bigger than $1-\alpha$ for any $\alpha > 0$ by choosing $n$ sufficiently large. \square\,

\end{document}